\documentclass[11pt]{elsarticle}
\usepackage{graphics} 
\usepackage{epsfig} 
\usepackage{amssymb}
\usepackage{amsthm}
\usepackage{amsmath}
\usepackage{verbatim}
\usepackage{xcolor}
\usepackage{siunitx}
\usepackage{textcomp}
\usepackage{lineno}
\usepackage[ruled]{algorithm2e}
\usepackage{subfigure}
\usepackage{hyperref}
\usepackage{caption}
\usepackage{mathtools}
\usepackage{bbm}
\usepackage{makecell}
\usepackage{multirow}
\usepackage{float}
\usepackage{algpseudocode}

\usepackage{rotating}
\usepackage{eqparbox}
\usepackage{multirow}
\usepackage{mathtools}

\journal{Ad Hoc Networks}
\begin{document}


\title{\textit{iTRPL}: An Intelligent and Trusted RPL Protocol based on Multi-Agent Reinforcement Learning}
\author{Debasmita Dey}

\author{Nirnay Ghosh\corref{cor1}}
\ead{nirnay@cs.iiests.ac.in}
\address{Department of CST, Indian Institute of Engineering Science and Technology, Shibpur, Howrah 711103, India}

\cortext[cor1]{Corresponding author}

\begin{abstract}
Routing Protocol for Low Power and Lossy Networks (RPL) is the de-facto routing standard in IoT networks. It enables nodes to collaborate and autonomously build ad-hoc networks modeled by tree-like destination-oriented direct acyclic graphs (DODAG). Despite its widespread usage in industry and healthcare domains, RPL is susceptible to insider attacks. Although the state-of-the-art RPL ensures that only authenticated nodes participate in DODAG, such hard security measures are still inadequate to prevent insider threats. This entails a need to integrate soft security mechanisms to support decision-making. This paper proposes \textit{iTRPL}, an intelligent and behavior-based framework that incorporates trust to segregate honest and malicious nodes within a DODAG. It also leverages multi-agent reinforcement learning (MARL) to make autonomous decisions concerning the DODAG. The framework enables a parent node to compute the trust for its child and decide if the latter can join the DODAG. It tracks the behavior of the child node, updates the trust, computes the rewards (or penalties), and shares with the root. The root aggregates the rewards/penalties of all nodes, computes the overall return, and decides via its $\epsilon$-Greedy MARL module if the DODAG will be retained or modified for the future. A simulation-based performance evaluation demonstrates that \textit{iTRPL} learns to make optimal decisions with time. 
\end{abstract}
\maketitle
\section{Introduction} \label{Intro}
The Internet of Things (IoT) connects physical devices to the Internet to form intelligent environments like smart homes, cities, industries, hospitals, etc. Networked IoT devices use the de facto routing protocol RPL~\cite{rfc6550} to build ad-hoc IPv6 networks and exchange messages with one another and the Internet. An RPL-enabled network is modeled by a tree-like structure called Destination-Oriented Direct Acyclic Graph (DODAG), where one of the IoT devices (or a gateway device) acts as a root node, and the rest are the non-root nodes. 

Though RPL ensures only legitimate nodes join a DODAG using its built-in authentication mechanism, it may be subjected to security threats posed by `insider attacks,' which are malicious attacks perpetrated on the DODAG by a node with authorized system access. The legitimate but malicious node(s) is/are either compromised by an external adversary or render(s) fault due to hardware failure. It can exhibit misbehaving features, such as deliberate packet drops, repeated connection refusals, corrupting the packets, flooding the network with spurious packets, etc. These misbehaving actions severely affect the network performance of DODAG and deny the other legitimate nodes due services. Thus, there is a need for regular monitoring of the nodes' behavior to thwart insider attacks.

Several methodologies using cryptographic techniques, geographic location, and modified RPL features have been proposed for securing the RPL~\cite{kamgueu2018survey}. None of the available methods were designed to thwart insider attacks. To this end, we believe that there is a need to integrate soft security mechanisms, such as trust, reputation, belief, etc., with the existing authentication scheme provided by the RPL to minimize the threats posed by insider attacks. Employing trust to secure RPL from malicious attacks has also been attempted recently \cite{9223748}. We note that such solutions have two main drawbacks. First, some trust-based solutions are not resource-friendly and cause depletion of network bandwidth and devices' residual energy \cite{al2023systematic}. Second, most work aims to find the best paths by choosing the trusted parent nodes, but direct hops between parent nodes may not be available for all paths. In such paths, other nodes can exhibit malicious behaviors~\cite{8998289}.

As mentioned above, RPL enables nodes to self-organize themselves into ad-hoc networks and carry out network operations. Since there is no centralized authority to manage the network, the participating nodes have to make decisions and take actions locally through collaborations. Therefore, there is a need to empower the DODAG nodes to compute and learn neighboring nodes' trust based on their behavior and later take actions through intelligent decision-making. We believe Reinforcement Learning (RL) can be used to support decision-making. RL projects state to action to maximize rewards. Given the ability to choose the correct action to maximize the reward, RL is ideal for situations where an agent must interact with the environment and perform sequential decision-making. 

In a DODAG, multiple parent nodes build the network without a central authority. As the network size is relatively small, nodes can share a common reward mechanism and collaborate to compute the optimal policy/strategy for the environment. The decentralized RPL DODAG scenario comprising several nodes is tailor-made for incorporating a variant of decentralized RL, called \textit{Multi-Agent Reinforcement Learning (MARL) with cooperative settings for homogeneous agents} \cite{zhang2021multi}. Below, we present a motivating example to elucidate our problem statement. 

\noindent \textbf{Motivating Example.} \begin{figure}
  \centering
  \includegraphics[scale = 0.45]{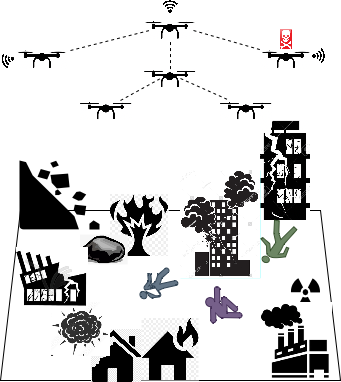}
  \caption{Motivating Example}
  \label{fig:mot}
  \vspace{-3mm}
\end{figure}
Unmanned Aerial Vehicles (UAV) have found implementations in many mission-critical applications where human involvement is challenging. One such scenario is a war zone, where it is dangerous for soldiers to inspect and collect data for an area under enemy attack. Suppose a group of UAVs belonging to an allied force arrives at a war zone to monitor the current situation. Let these UAVs use RPL protocol and organize themselves into a DODAG for message sharing and transferring to a remote control station. A situation may arise in which an adversary compromises one or more networked UAVs and makes them operate maliciously. It can result in dropping legitimate packets, refusing connection with peers, flooding the network with fake packets, changing data delivery paths by decreasing its own or increasing another node's rank, and so on. It is evident that even though all UAVs have legitimate identities, attacks could still originate and be organized within the network itself. Traditional authentication and authorization can't detect and prevent such anomalous behavior or insider attacks.      

Existing key-based authentication and authorization approaches cannot prevent insider attacks as they do not track a node's past purported behaviors or actions nor predict how the node will behave between the time the security keys were issued and used. No provision exists to determine if any node with a valid identity will cause any security problems. Addressing insider threats requires monitoring nodes' behavior and interrelations over a predefined time window. Hence, there is a need to incorporate \textit{soft security mechanisms} based on trust, risk, belief, etc., into the existing RPL. 

To address the challenge of mitigating insider attacks in RPL DODAGs, this paper proposes a framework called \textit{iTRPL} that attempts to strengthen the RPL protocol by incorporating a trust compute and provision model and multi-agent RL-driven decision-making. The significant contributions of the work are as follows:\\
\noindent(1) We implement trust computation and provisioning in RPL DODAG using (a) direct trust, which utilizes the node's behavior, and (b) indirect trust, which is obtained by combining trust opinions from neighboring nodes.\\
\noindent(2) We integrate the trust of the nodes with $\epsilon$-Greedy multi-agent reinforcement learning (MARL) to enable the root node of a DODAG to make decisions independently.\\
\noindent(3) We implement a custom simulation environment for performance analysis and validation.

The rest of the paper is organized as follows. Sec. \ref{pre} presents the preliminaries of RPL, trust, and MARL. Sec. \ref{sys_threat} describes the system model of the proposed framework. The details of the \textit{iTRPL} framework are provided in Sec. \ref{proposed}. In Sec. \ref{result}, we provide the implementation details and performance analysis. Sec. \ref{rel} reviews related works in RPL and implementation of trust-based RPL. Finally, we conclude and identify future work in Sec. \ref{conc}.

\section{Preliminaries} \label{pre}
This section presents basic details of the RPL protocol, trust computing and provisioning, and multi-agent reinforcement learning (MARL). 
\subsection{RPL Protocol} \label{rpl}
\begin{figure}
  \centering
  \includegraphics[scale = 0.3]{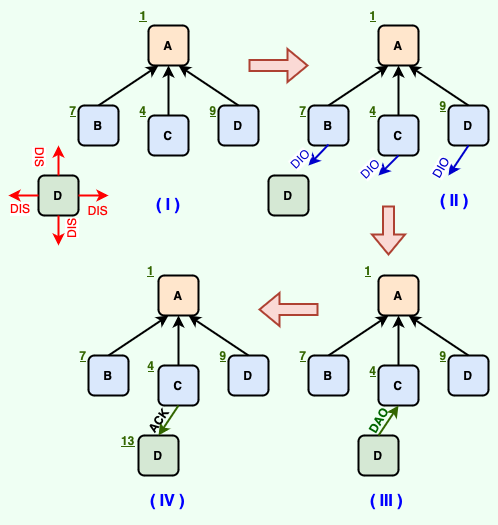}
  \caption{DODAG Formation}
  \label{dodag}
  \vspace{-5mm}
\end{figure}

RPL or IPv6 Routing Protocol for Low-Power and Lossy Networks works by building a Destination Oriented Direct Acyclic Graph (DODAG) \cite{rfc6550}. The DODAG consists of two types of nodes based on their ranks: parent and child. Predefined objective functions, like Objective Function Zero ($OF0$) and Minimum Rank with Hysteresis Objective Function ($MRHOF$) compute the ranks of the nodes \cite{lamaazi2020comprehensive} using metrics like hop count, residual energy, and estimated transmission count ($ETX$). The parent nodes are ranked higher than the child nodes. A DODAG operates using five types of control messages, namely, DIS (DODAG Information Solicitation), DIO (DODAG Information Object), DAO (DODAG Advertisement Object), DAO-ACK (DAO Acknowledgement), and CC (consistency check) messages. As shown in Fig. \ref{dodag}, if any node (child) wants to join a DODAG, it broadcasts DIS messages to obtain the ranks of parents. Each parent node responds with an unicast DIO message containing its rank. The child node selects a suitable parent based on rank and sends a DAO message to the latter. As show in Fig. \ref{dodag}, the new node sends DAO to the parent with the best rank (rank 4). The parent sends a DAO-ACK message as an acknowledgment after confirming the child node's inclusion in the DODAG. The CC messages are used to manage the network. Besides the control messages, the DODAG utilizes a trickle timer to maintain the frequency of broadcasting DIO messages. RPL supports three types of routing based on the direction of traffic \footnote{https://docs.contiki-ng.org/en/develop/doc/programming/RPL.html}: (1) \textit{Upward routing}: here data flows from any node to root node; (2) \textit{Downward routing}: data flowing occurs from root node to any node; (3) \textit{Any-to-any routing}: this case is more flexible, here data flows between any pair of nodes, and for carrying out the operation, both upward and downward routing is used. With respect to managing routing information, RPL operates in two modes: (1) \textit{Storing}: parent node stores information of its child node(s) and the latter sends DAO message to the former; (2) \textit{Non-Storing}: root node stores information of all other nodes and the DAO messages are sent directly to it.
\subsection{Trust} \label{sub:trust}
Trust is a socio-cognitive phenomenon that defines an association between two parties: \textit{trustor} and \textit{trustee}. The trustee is supposed to make itself believable to the trustor by taking trustworthy actions. In a computer network, a node is said to be trusted based on its performance in the network and the behavior exhibited towards its neighbors. Generally, trust is computed by a function that takes the evidence as input and is expressed as a score within a predefined interval. The different trust computing methods available in the literature are \cite{josang2007survey}: (i) \textit{Simple summation or average of ratings}: the overall trust score is obtained using the summation of the individual scores (positive or negative); (ii) \textit{Probabilistic systems}: they compute trust using probability density function which takes binary ratings as input; (iii) \textit{Discrete trust models}: here a trustee is labeled as trustworthy or untrustworthy at the beginning, then look-up tables or referrals are used to validate the label; (iv) \textit{Belief model}: here the trustee's trust is determined with the quantified value of specific parameters like belief, disbelief, and uncertainty; (v) \textit{Fuzzy models}: the membership functions determine the degree of trustworthiness; (vi) \textit{Flow models}: long or short chains of transitive relations determine the trust.

In a multi-agent environment, all agents may not have directly interacted with each other till a given time instance. In such a scenario, for an unknown trustee, a trustor has to depend on the personal experiences of other trustors who, in turn, have previously interacted with it. As evident, there exist two ways of provisioning trust scores \cite{parhizkar2020combining}:\\
\noindent \underline{(i) Direct trust}: In this method, the trustor assigns a trust score based on his/her personal experience following one or more interactions with the trustee.\\
\noindent \underline{(ii) Indirect trust}: In this method, a trustor depends on a trustee's reputation obtained through the latter's previous interactions with other peers. This reputation is built by combining positive and negative feedback of the peers given over time towards the trustee. 
\subsection{Multi-Agent Reinforcement Learning (MARL)} \label{marl}
Multi-Agent Reinforcement Learning (MARL), a variant of Reinforcement Learning, consists of multiple agents, and each agent is motivated by a reward resulting from the action it has undertaken according to the state of the environment. 
MARL can be categorized into three groups, namely \textit{fully cooperative}, \textit{fully competitive}, and \textit{combination of both} \cite{zhang2021multi}. In fully cooperative MARL, all agents are homogeneous and share a common reward function. They communicate with each other to optimize local/individual rewards to maximize the cumulative reward for the environment. One example of a cooperative MARL setting is the \textit{markov potential game}, where the common reward function is modeled as a potential game. In fully competitive MARL, the agents compete against one another to optimize individual rewards, which may result in diminishing returns as a reward of one agent is a loss of another. An example of fully competitive MARL is \textit{zero-sum Markov game}. In a combined MARL, the agents do not impose any restrictions on their relationship with each other. One example is the \textit{general sum game setting}, which consists of competitive teams with a few cooperative agents. Like RL, a MARL model has entities such as agents, state, environment, actions, policy, reward, return, and value function \cite{sutton2018reinforcement}. The concepts of ``exploration'' and ``exploitation'' are prevalent in RL/MARL for learning optimal returns or actions over time. For this, the agents follow a $\epsilon$-Greedy approach, where $\epsilon$ is the probability of choosing an action ($0 < \epsilon << 1$). In ``exploitation'', an agent greedily chooses the highest-valued action with probability $1 - \epsilon$ to increase its return immediately. In ``exploration'', the agent randomly chooses an action from the rest with probability $\epsilon$, expecting a long-term benefit in returns. 

\section{System Model}\label{sys_threat}
This section presents the system model for the proposed \textit{iTRPL} framework.
\subsection{System Model} \label{sys}
Our system model consists of networked IoT nodes that use the RPL protocol to self-organize themselves into a data network modeled as a tree-like destination-oriented direct acyclic graph (DODAG). In this work, we assume that the RPL DODAG operates in storing mode and supports only downward routing, where data transfer occurs strictly from parent to child node. The networked nodes may or may not be mobile. We also assume that the nodes are in a ``quasi-static'' state to ensure they are in close proximity to ensure building and maintaining a DODAG and using it for data delivery. 

We identify three scenarios under which the structure of an existing DODAG can change:\\
\noindent\underline{Scenario-1 (the DODAG wants a new node to join)}: In this case, any node from the DODAG broadcasts DIO messages consisting of the node's rank. Node that wants to join respond with DAO messages. On receiving the DAO message, the potential parent node sends a DAO-Ack message if it wants the node to join the DODAG as its child. In Fig. \ref{dodag}, this scenario occurs from steps (II) to (IV).\\
\noindent\underline{Scenario-2 (an external node wants to join the DODAG)}: In this case, the node broadcasts DIS messages. After receiving the DIS message, DODAG nodes, likely to be its potential parents, send DIO messages advertising their rank. The external node then sends a DAO message to the particular node selected as its parent. Joining is confirmed by a DAO-Ack message sent by the parent. In Fig. \ref{dodag}, this scenario is depicted from steps (I) to (IV).\\
\noindent\underline{Scenario-3 (a child node wants to change its parent)}: Here, a child node wants to change its parent and get attached to one with a better rank. It starts by sending DIS messages, followed by other control messages like DIO, DAO, and DAO-Ack.

Trust is important in determining the relation between the nodes in the DODAG. As mentioned in Sec. \ref{sub:trust}, the trust score for a DODAG node will be provisioned either by direct or indirect methods. For direct trust, the trustor node will explicitly interact with the trustee node, and if the latter exhibits ``misbehavior'', it will serve as evidence necessary for trust computation. We consider the following node actions as misbehaving instances: (i) frequent packet drops, (ii) significant delay in message forwarding, (iii) refusal of connection requests, and (iv) spurious packet generation. This work will leverage a probability density function-based approach for trust computation. More specifically, we will use the \textit{Inverse Gompertz (IG) function}\cite{gompertz1825xxiv} to model the variation of a node's trust with respect to the proportion of its misbehaving instances. The Inverse Gompertz output non-linearly decreases from its upper asymptote to reach the lowest value and reflects decay in the trust relationship in any multi-user environment~\cite{mousa2015trust}.

Indirect trust occurs if the trustor has no previous interaction experience with the trustee and depends on the trust feedback (computed as direct trust) from its peers who have interacted with the latter. The trustor combines these trust scores, giving more significance to the scores assigned in recent times. The node's local MARL model uses the trust score to support decision-making regarding DODAG security.    

Multi-Agent Reinforcement Learning (MARL) enables the nodes in a DODAG to decide independently which node should be allowed or denied to join the existing network based on trust score. In our work, the networked nodes attempt to achieve a common objective of securing the DODAG against insider attacks launched by legitimate but malicious nodes. To this end, we aim to use trust scores of the nodes to reward or penalize them and aggregate them to take actions. This motivated us to adopt the $\epsilon$-Greedy MARL with \textit{fully cooperative} setting, where all the agents are homogeneous and share a common reward function. Securing DODAG through mitigating insider attacks is a continuous process with repeated actions in a time frame, which we denote as epochs. Further, an epoch is delineated into multiple episodes to capture fine-grained operations. 

A typical $\epsilon$-Greedy MARL is defined as a tuple $M = \textlangle N, E, S, A, r, R,\\Q, P, \epsilon\textrangle$. We map the different entities in the $\epsilon$-Greedy MARL model to our context in the following way:\\
\noindent\underline{1. Agents ($N$)}: They refer to the IoT nodes that self-organize themselves into a DODAG without the support of any centralized authority. These agents are homogeneous regarding local resources and have a common reward function.\\
\noindent\underline{2. Environment ($E$)}: The environment in our context is the network of IoT nodes modeled as a single-root DODAG.\\
\noindent\underline{3. State ($S$)}: We consider two sets of states: (i)\textit{State of nodes} refers to the trust score of each of the nodes determined by its parent node; (ii)\textit{State of DODAG} refers to the return calculated by aggregating the latest rewards of the nodes in the DODAG in the current time epoch. \\
\noindent\underline{4. Action ($A$)}: In a MARL model, the joint actions of agents influence the state of the environment and the future rewards they receive. Our model has three actions: (i) \textit{allow} or \textit{deny} an agent (any node) to join an existing DODAG by a parent during an episode based on its trust score; (ii) \textit{change} of the parent by a node during an epoch; (iii) \textit{retain} or \textit{modify} the DODAG by the root agent (root node) at the end of an epoch based on the outcome of the state-action value function and $\epsilon$-greedy approach.\\
\noindent\underline{5. Reward ($r$)}: The reward provided in \textit{iTRPL} are at two levels: (i) assigned by the parent node to its children; (ii) assigned by the root node for weighing the state-action pairs.\\
\noindent\underline{6. Return ($R$)}: It is obtained at the DODAG root by aggregating other nodes' rewards at the end of an epoch.\\
\noindent\underline{7. State-action Q-value function ($Q$)}: The root uses this function to decide the outcome of a DODAG after each epoch based on the latter's state.\\
\noindent\underline{8. Policy ($P$)}: The MARL model has two sets of policies: (i) \textit{episode-level policy} that considers the trust score of a node and determines its reward; (ii) \textit{epoch-level policy} considers the values obtained from state-action Q-value function to determine the future outcome of the DODAG.\\
\noindent\underline{9. $\epsilon$-value}: The root node applies the $\epsilon$-Greedy approach on the results obtained from the state-action Q-values at the end of each epoch and finalizes its action.
\begin{figure*}[ht!]
  \centering
  \includegraphics[scale = 0.4]{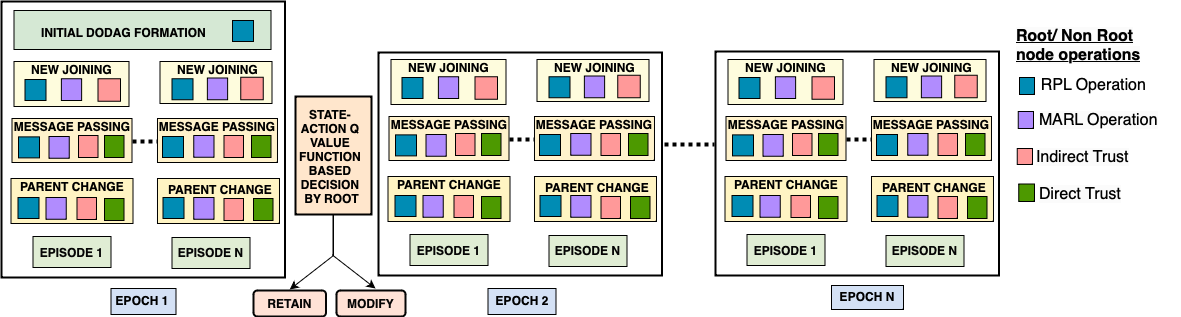}
  \caption{System Model}
  \label{sys_model}
\end{figure*}

\noindent Fig. \ref{sys_model} depicts our system model, which has five major components:\\
\noindent\underline{1. DODAG root node}: It denotes a special node in a DODAG that initiates the latter's origin and typically serves as a border router to the Internet. The root node's primary operations are (i) the inclusion of nodes at the first level of DODAG and acting as their parent through the episodes; (ii) aggregation of rewards of other nodes collected from their respective parents to compute an overall return; (iii) determine the state of the DODAG by comparing the return with the threshold return value; and (iv) calculate the state-action value function and apply the $\epsilon$-Greedy approach in order to decide on modifying or retaining the DODAG. In \textit{iTRPL}, we assume that the DODAG root is safe and cannot be compromised by any adversary.\\
\noindent\underline{2. DODAG non-root node}: A typical DODAG non-root node is characterized by the following attributes: \textit{node Id}, \textit{parent Id}, \textit{DODAG Id}, \textit{routing table}, \textit{failure rate}, \textit{rank}, \textit{ETX}, and \textit{trust score}. At any instance, the routing table of a particular node contains information like destination ID (one of its grandchildren), next-hop (one of its children), and trust score of children, etc. The routing table helps the node pass messages and keeps track of the trust score of its current and previous child nodes. The failure rate for a node is its intrinsic property and denotes the percentage of misbehaving instances (as perceived by its parent) exhibited over its lifetime. Depending on the failure rate, we classify the non-root nodes into three types: \textit{honest}, \textit{selfish}, and \textit{malicious}. A typical node's lifetime starts from joining the DODAG and continues until it leaves voluntarily or is suspended from the DODAG. The initial trust score for the new nodes (never been a part of the current DODAG) is set to 1.0 to avoid the ``cold-start problem.'' The trust score measures an existing node's reliability in performing DODAG network operations. It is updated by the parent it serves (direct trust) or the node that intends to add it as a child following a parent change action (indirect trust).\\ 
\noindent\underline{3. Episode}: An episode is a typical observation window in a MARL model where one or more operations can occur and repeat across other episodes. In \textit{iTRPL}, the following operations occur during a span of an episode: (a) computation of direct and indirect trust; (b) assign rewards to child nodes based on exhibited network activities (MARL operations); (c) new node allow/deny, parent change (RPL operations).\\
\noindent\underline{4. Epoch}: An epoch is a larger temporal window with a predefined number of episodes. At the end of an epoch, the action to be taken on the DODAG, whether it will be retained or modified based on the value function, is decided by the root node.\\
\noindent\underline{5. State-action Q-value function}: The state-action Q-value function will determine the quality of the DODAG and enable the decision of its existence. It is local to the DODAG root node and takes a pair of states and actions as input. The states are \textit{(High Return, Low Return)} denoting the aggregated rewards of the DODAG, and the pair \textit{(Retain, Modify)} indicates the action to be taken on the DODAG. The value function combines the states and actions and generates four state-action pairs to make a decision. 
\section{\textit{iTRPL} Framework} \label{proposed}
\begin{figure*}[ht!]
    \includegraphics[scale = 0.45]{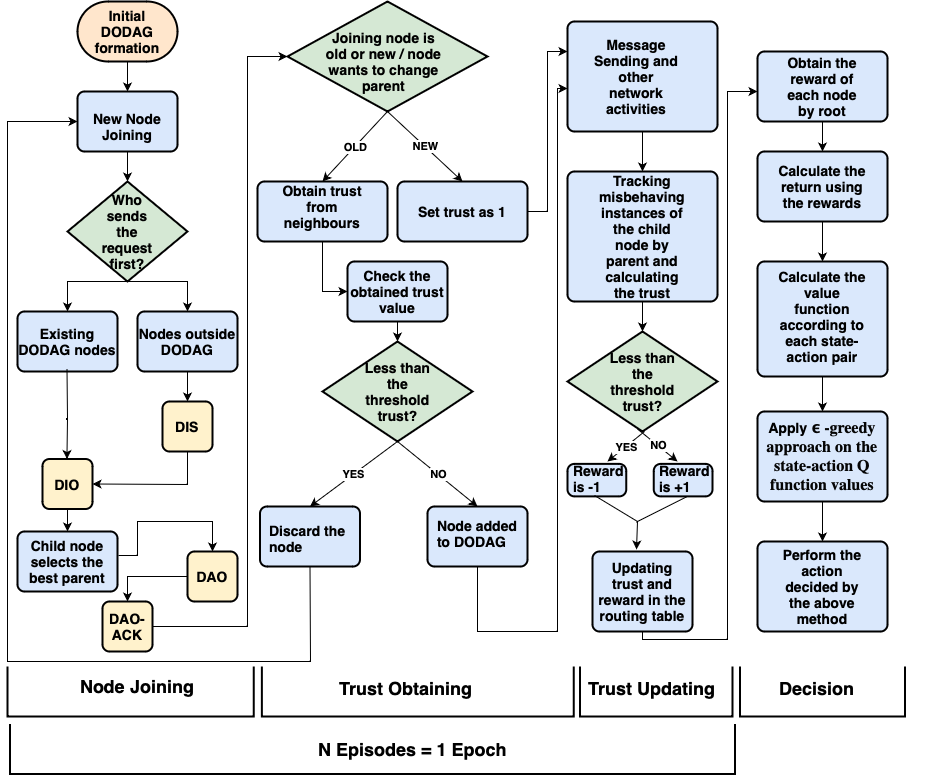}
    \caption{Flow Diagram for Various Operations in \textit{iTRPL}}
    \label{fig:overall}
\end{figure*}

This section presents the working principle of \textit{iTRPL} framework in detail. As described in Sec. \ref{sys}, the three main parts of the proposed approach are (i) DODAG node operations, (ii) computation and provisioning of node trust, and (iii) use of node trust-based reward in a $\epsilon$-Greedy MARL model to make decisions regarding the DODAG. For a better comprehension of the entire process, we present a flow diagram in Fig. \ref{fig:overall}. The following subsections elucidate these parts.
\subsection{DODAG Node Operations} \label{proposed:node}
A DODAG node has features like node ID, trust score, routing table, failure rate, etc. The node ID uniquely identifies a node in the DODAG. A node's trust score denotes its reputation from its parent's perspective. The local routing table for a node contains the following fields: (i) \textit{Destination}: denotes the destination node ID that this node would like to forward packets in the near future; (ii) \textit{Next hop}: it is the node id of the next node in the routing path, in order to reach the destination node; (iii) \textit{Trust}: the latest trust score assigned to the destination node through direct and/or indirect trust methods; (iv) \textit{Reward}: it denotes the latest reward obtained according to the policy (while calculating trust) by the destination node; (v) \textit{Episode}: the latest episode, in which the trust score and reward has been updated; (vi) \textit{Percentage of misbehaving instances}: the fraction of instances when misbehaving instances are exhibited by the destination node from the local node's perspective (across multiple episodes). 
\subsection{Computing Direct Trust} \label{trust_comp}
\begin{figure}[ht!]
\begin{center}
\subfigure[\label{Varying B}]{
\includegraphics[scale=0.25]{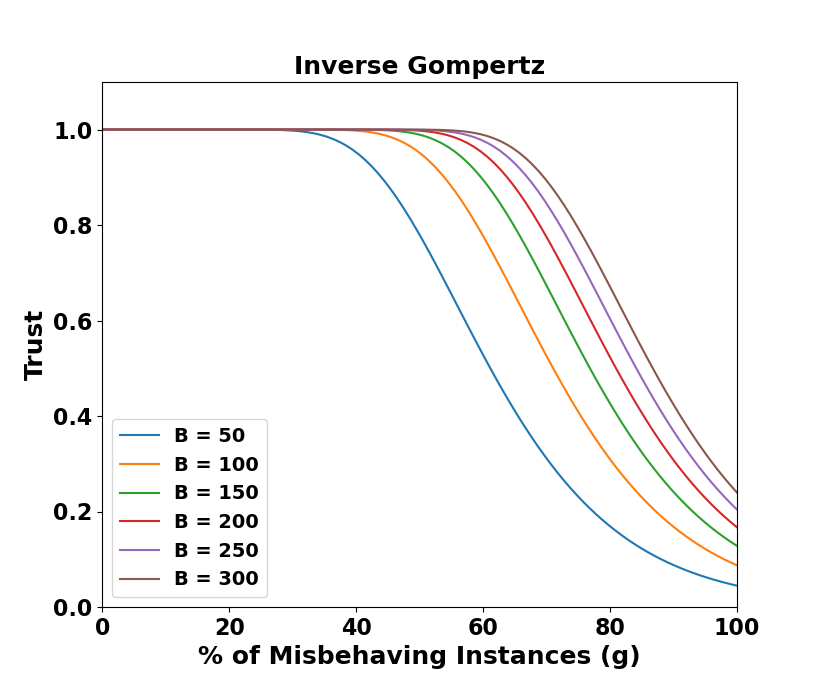}}
\subfigure[\label{Varying C}]{
\includegraphics[scale=0.25]{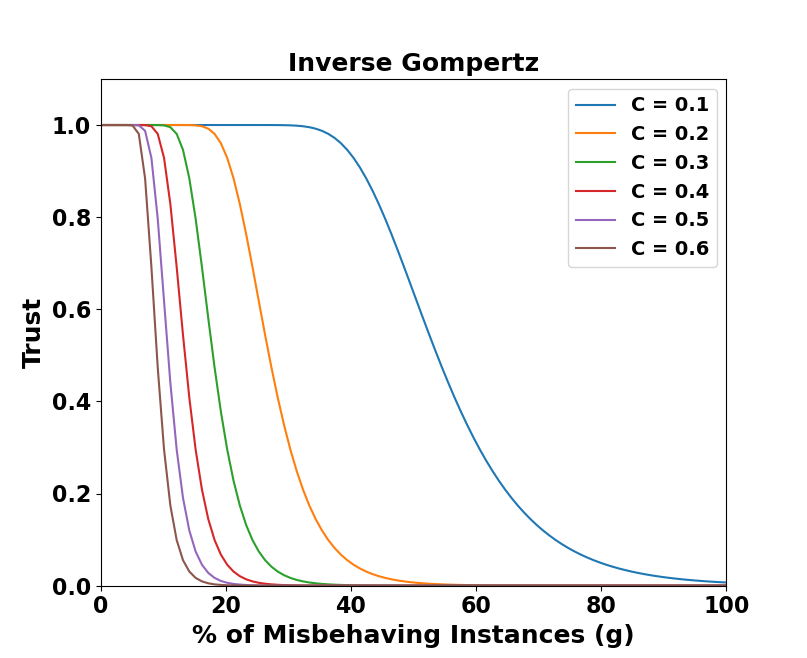}}
\caption{IG Function Parameter Study: (a) $B$ (b) $C$}
\label{Trust Obtained}
\end{center}
\vspace{-3mm}
\end{figure}
As discussed in Section. \ref{sys_model}, we use the \textit{Inverse Gompertz (IG)} function \cite{roy2020biosmartsense+} for computation of node trust. After a node joins under a parent in the DODAG, the latter observes the former's network activities over an episode. If the parent perceives any misbehaving instances, it uses the $IG$ function to update the node's trust. The mathematical form for the $IG$ function is given below:
\begin{equation}
     \tau = 1 - A . e^{-B.e^{-C.g}}
\end{equation}
where $\tau$ is the node's direct trust in the range [0, 1], $g$ is the percentage of misbehaving instances observed by the parent during an episode, $A$, $B$, and $C$ are the $IG$ function parameters to denote the initial asymptote, displacement of the trust along the X-axis, and decay parameter, respectively. We define the percentage of misbehaving instances as $g = \frac{MI\#}{NO\#}\times 100$, where $MI\#$ is the number of perceived misbehaving instances, and $NO\#$ is the number of network operations carried out by the node during an episode. 

It is to be noted that all new nodes join the DODAG with an initial trust of $\tau = 1.0$. At this stage, the percentage of misbehaving instances is nil. Thus, a new node is given the authority to carry out all network activities, believing it will not misuse its access privileges. This avoids the delay owing to the trust build-up phase and eliminates the cold-start problem essential for mission-critical applications. The parent node monitors the behaviors of the child nodes, and if it perceives any misbehaving instances, it uses the $IG$ function to penalize them by reducing the current trust score. At the end of each episode, the parent node updates its routing table with its child nodes' new trust scores.

Fig. \ref{Trust Obtained} illustrates the effect of $IG$ function parameters $B$ and $C$ on the trust score $\tau$.  As node trust is a real value in the interval [0, 1], we have fixed the value of the initial asymptote to $A = 1$. 
In Fig. \ref{Varying B}, we fixed the value of $C = 0.7$ and varied $B$ from 50 to 300. At a higher value of $B$, the trust score of the node is kept constant for a longer duration. Therefore, the parameter $B$ enables the parent node to exhibit tolerance towards child nodes. For instance, in a critical scenario, the parent may choose a low $B$ to drop a malicious child node's trust at a few misbehaving instances. It can set a higher $B$ for other cases, enabling the child node to continue its operations irrespective of misbehaving instances. 

In Fig. \ref{Varying C}, we fixed the value of $B = 150$ and varied $C$ from 0.1 to 0.6. Parameter $C$ is the decay parameter that controls the rate of decay of the trust scores, lower bound by 0. The trust score drops rapidly for higher values of $C$ with a relatively low percentage of misbehaving instances. The parent node can choose a smaller $C$ for a less critical application. In this situation, the parent node tolerates many misbehaving instances before the node trust drops significantly.

\subsection{Provisioning Indirect Trust} \label{pro:ind}
We calculate indirect trust under two circumstances:\\
(1) A new node seeks to join the DODAG; the parent node attempts to get knowledge of the node's trust value from the latter's previous parents.\\
(2) A child node seeks to change its current parent to another parent of a superior rank. Then, the new parent attempts to determine the trust of this child node.

We compute the indirect trust of any child node $k$, $\mathcal{T}_k$, using the following equation:
\begin{equation}
    \mathcal{T}_k =  \sum_{i=1}^{n} \omega_i \times \tau_i 
\end{equation}
Where $n$ denotes the number of previous parents of node $k$, $\tau_i$ is the direct trust computed by the $i^{th}$ parent, and $\omega_i$ is the corresponding trust coefficient that assigns a higher weight to recent episodes than older ones.   

The trust coefficient $\omega$ controls the effect of previous trust scores on the indirect trust of the node $k$ and its weight for any episode $t$ is calculated as \cite{6858002}:
\begin{equation}
    \omega = e^{-\alpha(T - t)}
\end{equation}
Here, $\alpha$ is a system parameter dependent on network condition, and $T$ is the total number of episodes. It is to be noted that all $T$ episodes belong to the same epoch. 
\begin{figure}[ht!]
\centering
    \includegraphics[width=35ex]{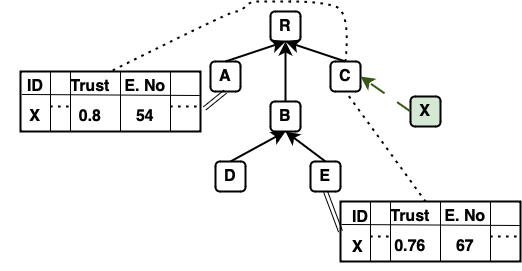}
    \caption{Provisioning of Indirect Trust}
    \label{fig:indirect_trust}
    \vspace{-3mm}
\end{figure}

Fig. \ref{fig:indirect_trust} showcases the indirect trust provisioning phenomenon for an arbitrary node $X$. Indirect trust provisioning can take place under two circumstances. First, node $X$ has left an existing DODAG at some previous episode. It seeks to rejoin the DODAG under a new parent, $C$. Second, node $X$ was previously a child of parent $E$ and now seeks $C$ as its new parent. In both cases, $C$ broadcasts \textit{CC} control messages asking for $X$'s trust score. As nodes $A$ and $E$ were $X$'s previous parents, they will provide trust values, say 0.8 as $\tau_1$ and 0.76 as $\tau_2$, respectively. Subsequently, the weights $\omega_1$ and $\omega_2$ will be calculated as $e^{-\alpha(T-54)}$ and $e^{-\alpha(T-67)}$, where 54 and 67 are the saved episode numbers in the routing tables of $A$ and $E$ in which the above-mentioned trust values have been saved. Suppose the total number of episodes $T$ is 70, and $\alpha$ is 0.09, therefore, $\omega_1$ will be 0.24 and $\omega_2$ will be 0.77. It is evident that $\omega_2$, which contains the latest episode number, will have more weight than $\omega_1$. The combined trust for node $X$,  $\mathcal{T}_X = (0.24 \times 0.8) + (0.77 \times 0.76) = 0.79$. With the increase in the total number of episodes, the calculated weights may become negligible or even negative. We will consider the corresponding $\omega = 0$ in that case. In a special case, if all $\omega$'s are 0, the node will be considered a new node with an initial trust value of 1. 

\subsection{MARL based Decision Making} \label{pro:marl}
In \textit{iTRPL}, we employ a \textit{fully cooperative} $\epsilon$-Greedy MARL model to support DODAG nodes in trust-based decision-making. Over several episodes in an epoch, the parent nodes assign individual rewards to all child nodes following the policy designed based on trust scores (episode-level policy). After the completion of an epoch, all parent nodes share the individual rewards of their child nodes with the DODAG root. The root uses these inputs to decide whether the DODAG should be retained for the next epoch or modified by deleting the node(s) with the lowest rewards. Below, we present the roles of root and other non-root nodes in MARL.
\subsubsection{MARL Operations at Non-Root Nodes} \label{pro:marl:nr}
In every episode, a non-root node observes the behavior of its child node $k$ and assigns a reward based on the following episode-level policies: 
\begin{itemize}
    \item If $\tau_k < \theta$, a new node is denied from joining the DODAG; for an existing node, it is rewarded with -1.
    \item If $\tau_k >= \theta$, a new node is allowed to join the DODAG, and for the existing node, it is rewarded +1.
    \item If a child node changes its parent, no reward is assigned.
\end{itemize}
Here, $\theta$ is a predefined trust threshold set by all parents. After all the episodes for an epoch have elapsed, the non-root node sends the latest reward for all its children to the root. 
\subsubsection{MARL Operations at Root Node} \label{pro:marl:r}
The root node holds the authority to decide if the DODAG will be retained or modified with the help of its state-action Q-value function and $\epsilon$-Greedy approach. Four state-action pairs comprising two actions \textit{Retain, Modify} and two DODAG states \textit{High Return, Low Return} are input to the Q-value function. \textit{High Return} is the aggregated reward of the DODAG nodes greater than a threshold return value. Whereas \textit{Low Return} is the sum of reward lesser than the threshold return. Since a parent can reward children only $+1$, $-1$, and $0$, the aggregated reward (return) should be in the range $[-N, +N]$, where $N$ is the number of non-root nodes in the DODAG. Unlike the threshold trust value, the threshold return is not fixed but is dynamically updated in each epoch, as the number of non-root nodes in the DODAG may change. We consider $\lceil \frac{N}{2} \rceil$ as the threshold for classifying high and low returns. The Q-value function uses the return to compute values for all four state-action pairs. The estimated value for a state-action pair $(s, a)$ following policy $\pi$ is calculated as \cite{sutton2018reinforcement}:
\begin{equation}
\begin{split}
    Q_*(s,a) & = max_{\pi *} Q_{\pi *}(s,a)\\
    & = max_{\pi *}E[G_{\iota} | S_{\iota} = s, A_{\iota} = a] \\
   &  = E[R_{\iota+1} + \gamma max_{a'}Q_*(S_{\iota+1},a')| S_{\iota} = s, A_{\iota} = a]\\
   & = \sum_{s', r} p(s', r|s, a)[r + \gamma max_{a'} Q_*(s', a')]
\end{split}  
\label{value_function}
\end{equation} 

In Eqn. (\ref{value_function}), the optimal state-action value function $Q_*(s, a)$ can also be expressed as $max_{\pi} Q_{\pi}(s, a)$ which is equal to the expected return for taking action $a$ in state $s$ as per an optimal policy $\pi$. Therefore, we can express the equation as $max_{\pi *}E[G_{\iota} | S_{\iota} = s, A_{\iota} = a]$, where $G_t$ is the return obtained by aggregating the rewards for each episode of epoch $\iota$. $G_{\iota}$ can be expressed as $E[R_{\iota+1} + \gamma max_{a'}Q_*(S_{\iota+1},a')| S_{\iota} = s, A_{\iota} = a]$, where $R_{\iota+1}$ is the expected return for the next epoch $\iota+1$, $S_{\iota}$ is the state in the current epoch $\iota$, and $S_{\iota+1}$ is the state in the next epoch. The above expression for expected return can be converted into \textit{Bellman's optimality equation} for $Q_*$ as $\sum_{s',r} p(s', r|s, a)[r + \gamma max_a Q_*(s', a')$ where $r$ is the expected reward, and $p(s', r|s, a)$ denotes the probability of obtaining reward $r$ for reaching state $s'$ from state $s$ by taking action $a$. Table \ref{tab:exp_rew} shows the expected rewards assigned to each state-action pair by the root node.
\begin{table}
\begin{centering}
    \caption{Expected State-Action Rewards}
    \label{tab:exp_rew}
    \begin{tabular}{|c|c|c|c|}
    \hline
    \textbf{Current State} & \textbf{Future State} & \textbf{Action} & \textbf{Expected Reward} \\
    \hline
    Low Return & High Return & Retain & +1 \\
    \hline
    High Return & Low Return & Retain & -1 \\
    \hline
    High Return & Low Return & Modify & +1 \\
    \hline
    Low Return & High Return & Modify & -1 \\
    \hline
    \end{tabular}
\end{centering}
\vspace{-5mm}
\end{table}

The term $\gamma max_a Q_*(s', a')$ is the discounted future state-action value, where $\gamma$ is the discount factor, and the sum is taken over all the possible next states $s'$ and rewards $r$. We calculate the optimal value $Q_*(s, a)$ for all four pairs: \textit{(High Return, Retain)}, \textit{(Low Return, Retain)}, \textit{(High Return, Modify)}, and \textit{(Low Return, Modify)}. The root node selects one of the four state-action pairs stochastically by following the $\epsilon$-Greedy approach.

It is to be noted that there are elements of uncertainty in our considered DODAG environment particularly with identifying a node's misbehaviors and trust computed on its basis. We depend on the parent node's subjective judgment to qualify its child node's network activities as normal or malicious. The action decisions taken by the root, based on the rewards computed from perceived node trust, will not be fair if taken deterministically. On the contrary, if the action decisions are taken stochastically, and the root learns optimal actions by alternating between `exploitation' and `exploration,' over several time epochs, it is likely to tilt the odds somewhat in favor of the DODAG nodes. Thus, we use the $\epsilon$-Greedy approach for probabilistically choosing the action (see Eqn. \ref{epsilon}). 
\begin{equation} \label{epsilon}
\mathbf {A} = \left\{
\begin{array}{l l}
  Max(Q_*(s,a)), & \quad \mbox{with\ probability\ (1-$\epsilon$)}\\\\
  a\ random\ action, & \quad \mbox{with\ probability\ $\epsilon$}\\
\end{array}\right.
\end{equation}
Eqn. \ref{epsilon} shows that for a given probability $\epsilon$, the root node either chooses the highest value state-action pair with a probability $1 - \epsilon$ (to utilize ``exploitation'') or selects any one of the other three state-action pairs randomly with a probability $\epsilon$ (to exert ``exploration'').  

If the chosen action is \textit{Retain}, the DODAG is retained with the existing node, and new nodes will be added in the next epoch. Otherwise, if the decision is \textit{Modify}, the root node attempts to alter the DODAG by suspending all nodes with $-1$ rewards (refer to Algorithm \ref{algo:modify}). The deleted non-root node may be a leaf or an intermediary node with children. Deletion of the leaf node is trivial. For the other case, the children of the node-to-be-deleted are made the children of the latter's parent, and the routing table information of the children is appended to the new parent's routing table. 
\begin{algorithm}
\begin{algorithmic}
\scriptsize
\renewcommand{\algorithmicrequire}{\textbf{Input:}}
\renewcommand{\algorithmicensure}{\textbf{Output:}}
\caption{\textsc{Algorithm to Modify DODAG}}
\label{algo:modify}
\Require $Current\ DODAG$
\Ensure $Modified\ DODAG$
\State $Reward\_List[\ ] \gets [node\_list]_{Reward}$\\
\While{$Reward\_List[node] == -1$}
{\If {$children(Node\_to\_Delete) == True$}
{\State $children\_list \gets extract\_child(Node\_to\_Delete)$
\State $parent(children\_list) = parent(Node\_to\_Delete)$
\State $r\_table_{parent(children\_list)} \gets r\_table_{parent(children\_list)} \cup r\_table_{children\_list}$\Comment{Routing table updated}}
\State $delete(Node\_to\_Delete)$
\State $Reward\_List \gets Reward\_List \setminus (Node\_to\_Delete)_{Reward}$}
\end{algorithmic}
\end{algorithm}

\section{Results and Discussion} \label{result}
In this section, we evaluate the performance of the \textit{iTRPL} framework by validating actions taken based on the trust score generated by the framework. We developed a prototype of the framework in Python and simulated an environment consisting of several IoT nodes that use the RPL protocol to build and manage a network modeled as a DODAG~\footnote{https://github.com/debasmitadey9/iTRPL.git}. The details of the implementation and simulation environment are given below.
\subsection{Implementation Details}\label{imp_det}
This section presents the implementation details of the custom simulation environment. 
\subsubsection{DODAG Formation and Messaging}\label{imp:dodag} At the beginning, we create a root node and assign it a rank of 1. The other built-in characteristics of the root are $Id = 1$, failure rate as 0.05, trust as 1, and parent ID is set to NULL. It creates its routing table using its rank and the DODAG version number, and the rest of the fields are initially set to NULL. Next, the root node starts the function called \textit{trickle\_timer} and invokes a custom \textit{broadcast\_DIO} function, inviting other nodes to join the DODAG \cite{rfc6550}. Then, a trail of control message exchanges with functions \textit{send\_dao}, \textit{receive\_dao}, and \textit{send\_dao\_ack} takes place. These message exchanges also occur for non-root nodes, where the DODAG nodes want new nodes to join. If a new node attempts to join the DODAG, it starts with a \textit{send\_DIS} function.

We create multiple non-root nodes of the DODAG with consecutive node IDs, empty routing tables, DODAG version ID set to null, and rank set to 0. These nodes receive the DIO messages through a custom function \textit{receive\_DIO}. Message-sending operations occur using function \textit{send\_message} in between episodes where the following operations take place periodically: (i) choose source and destination nodes; (ii) create consistency check (CC) message (having fields \textit{RPLInstaceID}, \textit{Request/Response}, \textit{flag}, \textit{CC\ Nonce}, \textit{DODAGID}, \textit{Destination\ Counter}); (iii) look up the routing table of the source node for getting the next hop; (iv) generate misbehaving instances.
\subsubsection{Rank of Nodes}\label{imp:rank} For calculating the rank of a new node, the parent nodes use objective function $OF0$, which considers parameters such as the $ETX$ values and hop counts from the parent to the child node~\footnote{https://www.rfc-editor.org/rfc/rfc6552.html}. We calculate \textit{rank\_Inc} using the hop count between the probable parent node and child node, the $ETX$ of the child node, and a constant, called the \textit{rank\_factor}. The \textit{rank\_Inc} is added to the rank of the preferred parent to obtain the rank of the child node.
\subsubsection{Parent Selection}\label{imp:par} After receiving DIO message(s) from one or more parent node(s), the potential child node obtains its/their rank(s) from the message(s) and finds the one that has the best rank (if it received multiple DIO requests). It sends a DAO message using function \textit{send\_DAO}, confirming its join.
\subsubsection{Types of Nodes}\label{imp:nodes} Nodes and their behaviors are central to the proposed trust-based DODAG security. We have divided the non-root nodes based on their failure rates into the following categories:\\
(a) \underline{Honest}: These nodes are well-performing and rarely indulge in malicious activities. The failure rate percentage assigned to these nodes is $0\%$ to $10\%$. \\
(b) \underline{Selfish}: These nodes exhibit on-off behavior and are hard to identify. At times, they behave maliciously, while at other times, they carry out network activities normally. The percentage of failure assigned is $40\%$ to $50\%$.\\
(c) \underline{Malicious}: This kind of node misbehaves at a higher frequency and aims to degrade the overall DODAG performance. The failure percentage assigned is $80\%$ to $90\%$.
\subsubsection{Implementation of Failures}\label{imp:fail} Every node has an intrinsic failure rate. Parent nodes keep track of the instances of misbehavior to calculate the trust scores for the child nodes. In our case, we have considered packet drops, repeated connection refusals, corrupting the packets, and flooding the network with spurious packets to be misbehaving activities. We treat all misbehaving instances equally and use the node's failure rate as the likelihood of exhibiting one or more misbehaving instances. 
\subsubsection{Implementation of MARL and Trust related Operations}\label{imp:marl}
As stated in Sec. \ref{pro:marl}, MARL operations in \textit{iTRPL} start with the predefined trust-related policy. In every episode, the function \textit{obtain\_trust} returns the indirect trust value of the new node. Direct trust is calculated using functions \textit{compute\_trust} and \textit{inv\_gompertz}. Based on the trust, the action taken by a parent is ``allow'' or ``deny'' joining the DODAG. The parent refers to the \textit{policy} for rewards and then uses the \textit{compute\_rewards} function for assignment. At the end of an epoch, the root node calls the \textit{calculate\_value\_function} to compute the values of four state-action pairs. Next, the root calls the function \textit{epsilon\_greedy} and chooses the action. For the ``modify'' action, the root calls the \textit{modify} function (implementation for Algorithm \ref{algo:modify}) to suspend node(s) with $-1$ reward. In the process, re-assigning new parent(s) and transferring routing tables is done by the \textit{adjust\_routing\_table} function.
\begin{table}[ht!]
    \centering
    \caption{Parameters and Their Values}
    \begin{tabular}{|c|c|}
    \hline
        \textbf{Parameters} & \textbf{Values} \\
        \hline
         $\alpha$ (Parameter for $\omega$) & 0.05 \\
         \hline
         $\gamma$ (Parameter for $Q_*(s,a)$) & 0.8 \\
         \hline
         $A$ (Initial asymptote in IG function) & 1.0\\
         \hline
         $B$ (Displacement parameter in IG function) & 150 \\
         \hline
         $C$ (Decay parameter in IG function) & 0.7 \\
         \hline
         $\theta$ (Trust threshold) & 0.5 \\
         \hline
         No. of episodes per epoch & 10 \\
         \hline
         No. of epochs & 140\\
         \hline
         $\epsilon$ ($\epsilon$-Greedy approach) & 0.2 \\
         \hline
    \end{tabular}
    \label{tab:parameter}
\end{table}
\subsubsection{Simulation Environment}\label{imp:sim_env}
We have considered the following types of environments with some non-root nodes and one root node: (i) less malicious environment where the DODAG has $10\%$ malicious non-root nodes; (ii) medium malicious environment where the DODAG is formed with $40\%$ to $45\%$ of malicious non-root nodes; (iii) highly malicious environment where $85\%$ to $90\%$ non-root nodes are malicious. The values of different parameters used in our simulation are presented in Table~\ref{tab:parameter}.
\subsection{Performance Analysis} \label{performance}
This section provides a detailed performance analysis of the \textit{iTRPL} framework in terms of the quality of the decisions taken at different stages of the life-cycle of a DODAG.
\subsubsection{Study of $\epsilon$-Greedy Approach for Selecting Optimal Actions}
\begin{figure}[ht!]
\centering
    \includegraphics[scale = 0.3]{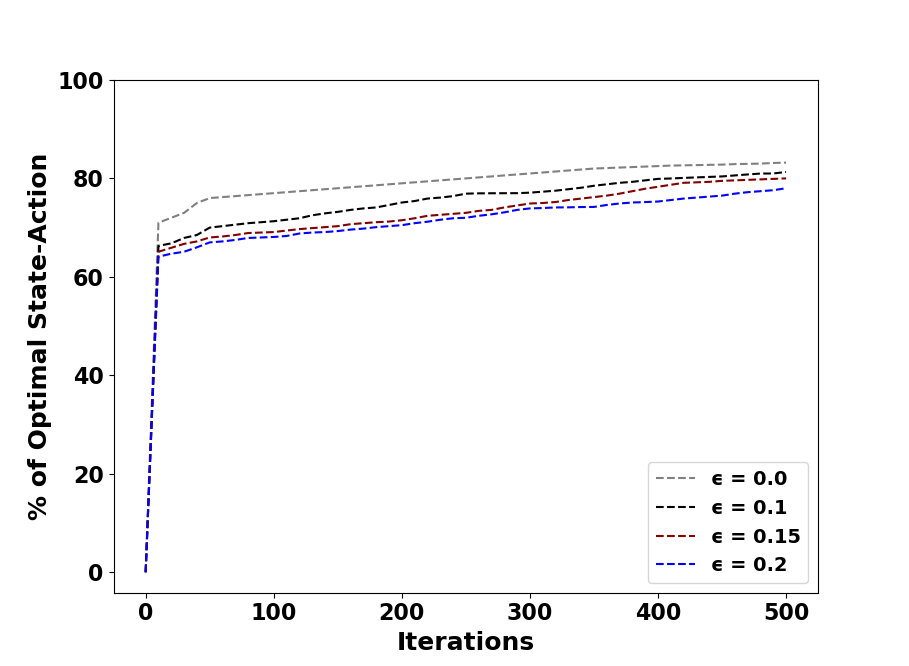}
    \caption{Effect of $\epsilon$ values on Optimal Action}
    \label{fig:epsilon}
    \vspace{-3mm}
\end{figure}

In \textit{iTRPL}, the state of the DODAG gets continuously modified owing to the inclusion and elimination of the nodes at the end of every epoch. Therefore, the return value of the DODAG can not be treated as the entity of learning by the MARL module. On the contrary, the MARL module in the root should learn to make the optimal decisions of retaining or modifying the DODAG over time. In Fig. \ref{fig:epsilon}, we conducted an empirical study to obtain the percentage optimal state-action pairs \textit{High Return, Retain} and \textit{Low Return, Modify} under various values of $\epsilon$, observed for 500 time-steps. At $\epsilon = 0$, MARL takes actions deterministically, which yields the highest percentage of selecting the best optimal state-action pair. However, in dynamic and uncertain environments, taking deterministic actions is unfair. With a gradual increase in the $\epsilon$ value, the root chooses increasing proportions of sub-optimal actions but converges over time, suggesting that it eventually learns to take optimal actions consistently. 

For simulation-based experiments, we have set $\epsilon = 0.2$  to ensure $20 \%$ times the root node adopts ``exploration'' by randomly choosing any of the actions resulting from the other three state-action pairs and $80 \%$ of the time it will ``exploit'' the action given in the highest-valued state-action pair. 
\subsubsection{Variation of Node Trust} 
\begin{figure}[ht!]
  \centering
  \includegraphics[scale = 0.27]{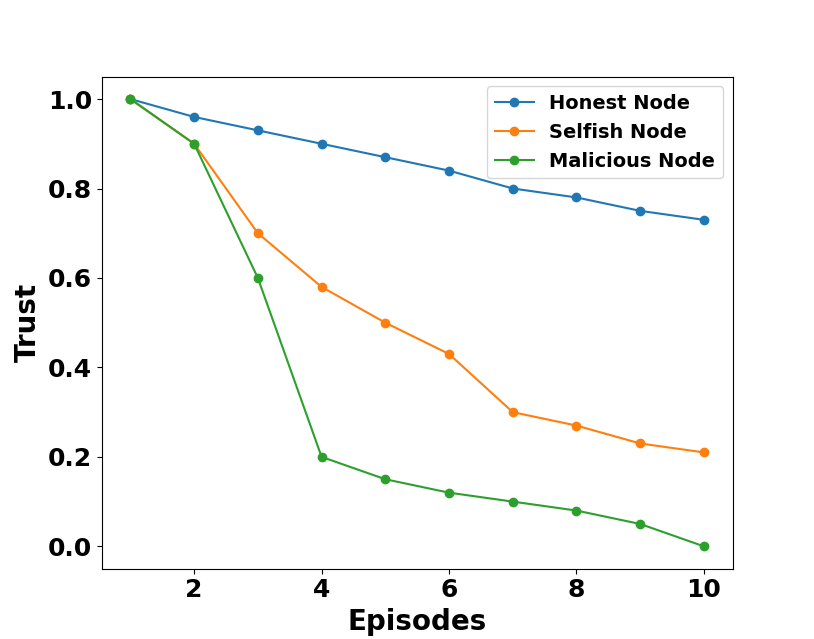}
  \caption{Node Trust over Episodes}
  \label{trust}
  \vspace{-5mm}
\end{figure}

Fig. \ref{trust} represents the variation of trust score for honest, selfish, and malicious nodes over ten episodes across all simulation environments. The trust depletion for honest nodes is significantly less than for selfish and malicious nodes. 
\subsubsection{Variation of Failure Rates} 
\begin{figure}[ht!]
  \centering
  \includegraphics[scale = 0.27]{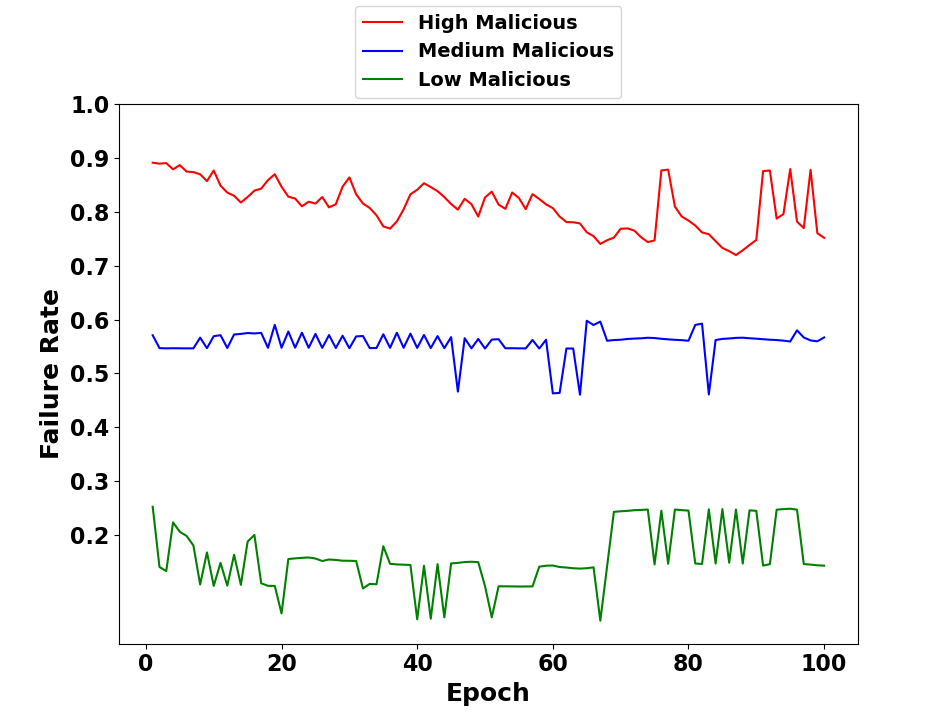}
  \caption{Failure Rates of Nodes over Epochs}
  \label{pdr}
  \vspace{-4mm}
\end{figure}

Fig. \ref{pdr} showcases the average failure rates exhibited by DODAG nodes over 100 epochs in three simulation scenarios: (i) DODAG contains a majority of malicious nodes; (ii) DODAG contains a fairly equal number of honest and malicious nodes; (iii) DODAG containing less malicious nodes than the honest ones. 
\subsubsection{Validation of DODAG-related Decisions}  
\begin{figure}[ht!]
  \centering
  \includegraphics[scale = 0.4]{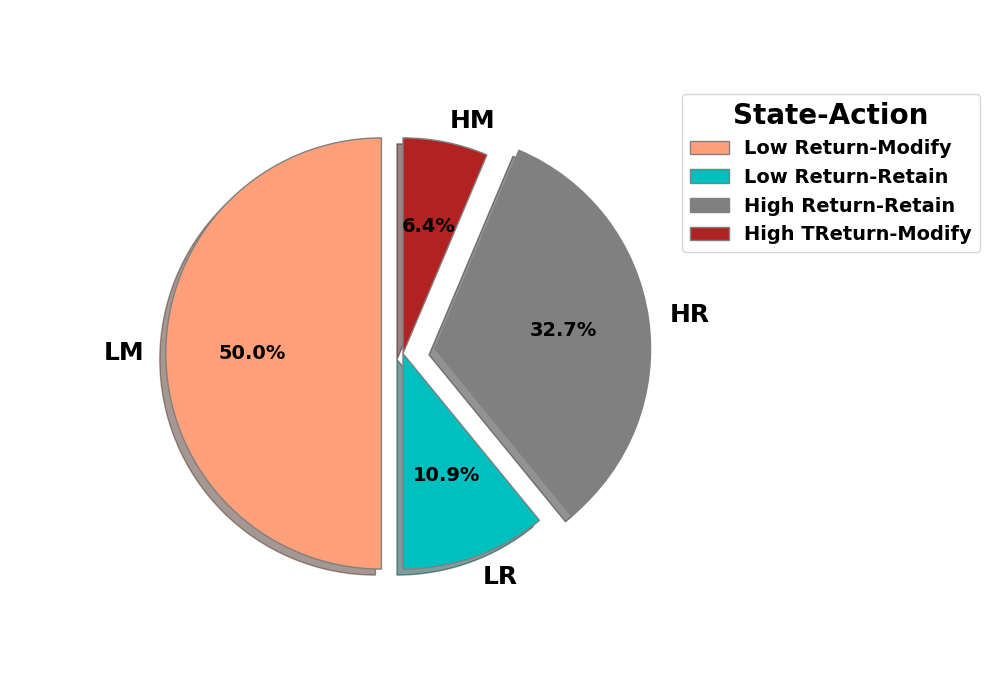}
  \vspace{-8mm}
  \caption{Decisions based on State-Action Pairs and $\epsilon$-Greedy Approach}
  \label{Decision}
  \vspace{-6mm}
\end{figure}

Fig. \ref{Decision} shows the distribution of action decisions taken by the MARL module in the root node over several epochs. As stated in Sec. \ref{pro:marl:r}, the native MARL module uses the Q-value function to generate values for the four state-action pairs and an $\epsilon$-Greedy approach to decide the action that needs to be taken. Intuitively, the two most desirable actions are (i) modifying the DODAG under low return ($LM$) and (ii) retaining the DODAG under high return ($HR$). This is reflected in the experimental study as these two state-action pairs have occurred in higher proportions (around 82.7\% of times). This implies that the MARL enables the root to learn optimal actions over time. As the $\epsilon$-Greedy approach selects actions stochastically, a few incorrect action decisions have occurred when state-action pairs $HM$ and $LR$ are selected (around 17.3\% of times). The percentage of selection of sub-optimal actions is expected to reduce further if the root carries on with the $\epsilon$-Greedy approach for more epochs.  
\subsubsection{Effect of Environment on DODAG-related Decisions} 
\begin{figure}[ht!]
  \centering
  \includegraphics[scale = 0.25]{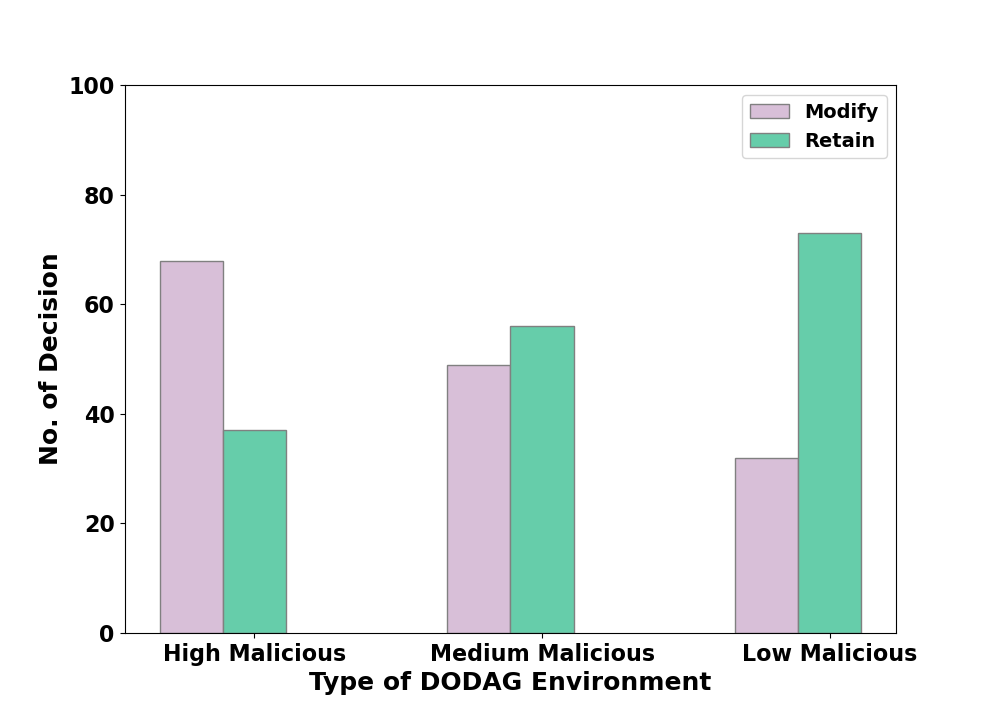}
  \caption{Decisions based on the Environment}
  \label{nodes}
  \vspace{-4mm}
\end{figure}

As stated in Sec. \ref{imp:sim_env}, we consider three environments based on the distribution of malicious nodes. Fig. \ref{nodes} shows the decisions taken under the three scenarios. It is evident that if the DODAG environment is highly malicious, it is modified a maximum number of times. On the contrary, in a low malicious environment, the number of times it is retained is more than the number of modifies. In a moderately malicious DODAG environment, modification and retention are comparable. Therefore, the proposed \textit{iTRPL} framework makes optimal decisions most often under varying degrees of malicious nodes in the DODAG. 
\subsubsection{Return Values in Different Simulation Environment}
\begin{figure}[ht!]
  \centering
  \includegraphics[scale = 0.3]{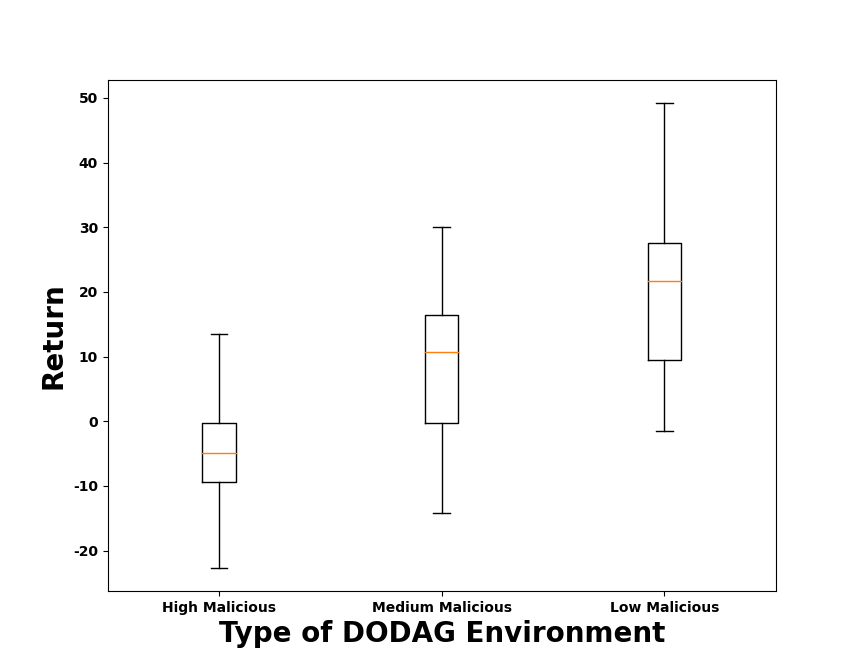}
  \caption{Return Values in Different Simulation Environment}
  \vspace{-3mm}
  \label{return}
\end{figure}

Fig. \ref{return} shows the ranges and means of return value computed by the root node for the existing DODAG in three scenarios. We observe that in a highly malicious environment, the mean return value is lowest, as the trust scores of non-root nodes will be less, resulting in low rewards across all episodes in an epoch. In contrast, the return value for a medium malicious environment is slightly higher, and the return of the less malicious scenario is highest.

\section{Related Works} \label{rel}
This section presents a brief literature survey about various approaches towards securing the RPL protocol from insider attacks.
\subsection{Mitigating Insider Attacks in RPL}
The use of RPL has been seen extensively in different IoT applications, such as healthcare \cite{refaee2022secure}, smart environments \cite{9709808}, transport \cite{SHARMA202312}, industry \cite{aydogan2019central}, military \cite{kannimuthu2021decision}, etc. The protocol successfully transfers packets from one resource-constrained node to another, maintaining bidirectional connectivity, flexibility, and robustness. Though the performance is commendable, there have been reports of several insider attacks in RPL. Insider attacks are generally stealthy and disrupt the network by bypassing RPL security mechanisms. The authors in \cite{bang2023impact} discuss ``rank attack'' in RPL, which changes its regular topology to degrade the node's QoS and energy. In \cite{sharma2023performance}, authors discuss a ``version attack'' that modifies the DODAG version number and makes it unstable and susceptible to Denial of Service (DoS) attacks. The authors in \cite{pu2020sybil} point out that owing to the Sybil attack in the RPL network, the trickle timer is restarted multiple times, leading to the rapid depletion of the nodes' energy. Other insider attacks that jeopardize a DODAG network are the replay attack \cite{bang2023impact}, flooding attack \cite{rouissat2022potential}, selective forwarding attack \cite{dhingra2022study}, and wormhole attack \cite{sharma2022analysis}. In \cite{alsukayti2022lightweight}, the authors modify the DIO and DAO messages format to dodge attacks. A Contiki-OS-based simulation environment has been developed to claim the efficacy of their approach. However, changes in the existing RPL protocol may not be compatible for implementation across all platforms. The technique adopted for mitigating rank attack by \cite{bang2022embof} changes the RPL objective function and proposes a rank check algorithm by a central entity. The authors present an elaborate performance analysis to validate their approach. Issues of scalability and single-point failure persist due to its centralized nature.

\subsection{Trust-based Approaches to Secure RPL}
The authors in \cite{8651846} propose a trust-based approach for RPL, where the nodes select the best path for data transmission using a trust value. The parameters for calculating trust value are nonce ID, timestamp, and network whitelist table. The mechanism mostly depends upon received signal strength, which, if not appropriately received, may not allow the protocol to work correctly. Behavioral trust has also been applied as a security measure in RPL. Behavioral trust can be used in \cite{azzedin2023mitigating}, where the neighboring nodes determine the trust value to secure the RPL protocol against DoS attacks. In \cite{kim2022physical}, the authors utilize trust calculated on the physical characteristics of the nodes to identify the best path for routing. We observe that trust-based solutions work on choosing the best parent but ignore the optimal path simultaneously. Moreover, parents are chosen at certain stages. Still, in other instances, when the nodes attempt to exchange messages (network statistics, routing information, node condition, etc.), the nodes may start misbehaving, but these instances are ignored. 

\subsection{Use of ML/RL in Securing and Improving RPL} 
In \cite{osman2021ml}, the authors have designed a data set to prevent version attacks on RPL and applied the Light Gradient Boosting Machine (LGBM) to detect irregularities in the data. In order to find out the best parent and hence the optimal route from one node to another, the authors in \cite{santos2023ml} trained a Gradient Boosted Decision Tree model using metrics from the routing table, such as neighboring nodes, next hop, a destination address, etc. In \cite{farag2021congestion}, the authors use Q-Learning for optimal parent selection. Similarly, \cite{golla2022efficient} also adopted the best parent selection with Q-Learning. Other work like \cite{saleem2020intelligent} embarks on learning automata for node learning and strengthening the adaptive nature of the RPL environment.

Most ML/RL-based solutions concentrate on attack detection, finding the best parent, searching for the best path, etc. If a large fraction of nodes in a DODAG is compromised, these approaches fail to render solutions. In contrast, \textit{iTRPL} ensures that the existing DODAG nodes allow only trusted nodes to join the DODAG and take preventive actions anytime during the life cycle of the DODAG by learning from the environment. Unlike other ML/RL approaches, MARL-based \textit{iTRPL} not only detects malicious nodes but also takes action to remove them from the DODAG.

As evident, most of the existing solutions are centralized and attempt to modify the original RPL protocol specifications. This is not conducive for emergencies where the nodes, without any centralized authority, have to self-organize themselves and ensure the security of DODAG. Additionally, changing the protocol specifications may cause compatibility issues. These solutions are applicable during the formation phase of DODAGs. No prevention is proposed for attacks occurring at intermediate stages.  
\section{Conclusion} \label{conc}
RPL is the widely used routing protocol for low-power and lossy networks that organizes the networked nodes in the form of a DODAG. In spite of built-in authentication mechanism, the protocol is susceptible to insider attacks. Such attacks are difficult to detect and prevent by traditional hard security mechanisms like authentication, access control, and identity management. Hence, soft security mechanisms like trust are required to mitigate threats owing to insider attacks. To mitigate threats due to insider attacks in RPL, this work proposed an intelligent trust-based framework, \textit{iTRPL}, that manages the trust scores of the DODAG nodes based on perceived misbehaving instances. A $\epsilon$-Greedy MARL model running at the DODAG root collaborates with other non-root nodes to make trust-based decisions. The actions corresponding to the MARL decisions are either to retain or modify the DODAG and are taken stochastically. The validity of the decisions is established through extensive simulation-based performance analysis. In the future, we plan to extend the scope of the environment by including multiple connected DODAGs having multiple roots as deployed in real-world scenarios and study the performance.
\bibliographystyle{unsrt}
\bibliography{biblio}
\end{document}